\documentclass[sigconf,nonacm]{acmart}

\usepackage{rotating}
\usepackage{colortbl}
\usepackage{multirow}

\begin{document}

\title{Deriving Semantics-Aware Fuzzers from Web API Schemas}

\author{Zac Hatfield-Dodds}
\affiliation{
    \institution{Australian National University}
    \city{Canberra}
    \country{Australia}
}
\email{zac.hatfield.dodds@anu.edu.au}

\author{Dmitry Dygalo}
\affiliation{
    \institution{schemathesis.io}
    \city{Prague}
    \country{Czech Republic}
}
\email{dmitry@dygalo.dev}

\begin{abstract}

Fuzzing---whether generating or mutating inputs---has found many bugs and security
vulnerabilities in a wide range of domains.  Stateful and highly structured web APIs
present significant challenges to traditional fuzzing techniques, as execution feedback
is usually limited to a response code instead of code coverage and vulnerabilities
of interest include silent information-disclosure in addition to explicit errors.

Our tool, Schemathesis, derives structure- and semantics-aware fuzzers
from web API schemas in the OpenAPI or GraphQL formats, using
property-based testing tools \cite{MacIver2019}. Derived fuzzers
can be incorporated into unit-test suites or run directly, with or
without end-user customisation of data generation and semantic checks.

We construct the most comprehensive evaluation of web API fuzzers to date, running
eight fuzzers against sixteen real-world open source web services.
OpenAPI schemas found in the wild have a long tail of rare features
and complex structures.  Of the tools we evaluated, Schemathesis was the
only one to handle more than two-thirds of our target services without a fatal
internal error.  Schemathesis finds $1.4\times$ to $4.5\times$ more unique defects
than the respectively second-best fuzzer for each target, and is the
\emph{only} fuzzer to find defects in four targets.

\end{abstract}

\maketitle

\section{Introduction}\label{introduction}

Much modern software communicates over the internet, and each service
therefore defines some kind of web API - often via a REST \cite{REST}
or more recently GraphQL \cite{GraphQL} architecture. Many services
also provide machine-readable schemas or specifications which describe
their input and output contracts and their semantics within the REST or
GraphQL architecture.

Such API schemas can be used to derive a \emph{service-specific} fuzzer that
can use the known structure of valid inputs and sequences of actions to
focus effort on interesting parts of an otherwise intractably large
search space. Derived fuzzers can also check for otherwise silent semantic
errors such as non-conforming responses, missing headers, or silent
information-disclosure vulnerabilities, based on standard schema and
HTTP semantics.

In practice OpenAPI \cite{OpenAPI} or GraphQL schemas are often
\emph{unsound}, permitting inputs or actions not handled by the
service. This may be due to errors in the schema or the service implementation,
because working with sound schemas is uneconomical, or
because of application-level constraints---such as database constraints,
order of event timestamps, relations between endpoints, etc.---which
cannot be expressed in the schema. We therefore see an
ongoing role for human judgement in customising automatically derived
fuzzers, analysing failures, detecting over-restrictive schemas, and
hand-coding additional or more precise tests.

\vfill\null \vfill\null  
\noindent
The main contributions of this paper are to:
\begin{itemize}
\item
  describe Schemathesis, a property-based testing library to
  automatically derive customisable web API fuzzers from OpenAPI or
  GraphQL schemas;
\item
  demonstrate that Schemathesis discovers more defects and handles more
  real schemas than previous tools;
\ifdefined\ABLATION
\item
  conduct an ablation study to investigate which features are
  responsible for Schemathesis' outperformance;
\else\fi and
\item
  provide a large evaluation suite of real web services, schemas, and
  fuzzers as reusable containers for use in future research.
\end{itemize}

\subsection{Property-based testing}

Property-based testing (PBT) originated with the Haskell library
QuickCheck \cite{QuickCheck}, which emphasised testing algebraic
properties of functions by generating many random inputs to a test
function. PBT generally differs from fuzzing more in workflow and the
affordances of tooling than fundamental concept (typically focussing
on highly-structured and always-valid data, along with integrated
shrinking). The definitive PBT library for Python is Hypothesis
\cite{MacIver2019}, with an estimated five hundred thousand users and
dozens of third-party extensions. Hypothesis is explicitly designed as a
library of tools to construct fuzzable tests---which can detect errors
that direct fuzzing would not---and a non-user-visible
bytestring-oriented fuzzing backend for them\footnote{\url{https://hypothesis.works/articles/what-is-property-based-testing/}}.

This abstraction has been very successful. The Hypothesis backend
includes integrated shrinking and error-deduplication
\cite{MacIver2020}, targeted property-based testing
\cite{TargetedPBT,AutomatingTargetedPBT,IJON}, swarm
testing \cite{SwarmTesting}, and optionally coverage-guided
fuzzing. Structured inputs are
defined using parser-combinators (`strategies'\footnote{
  Most PBT libraries call their description of possible values ``generators''.
  Because generators are a builtin type in Python, Hypothesis instead
  names them ``strategies''.});
and only the primitives supplied by Hypothesis have any knowledge of the
backend. Such separation of concerns has made it practical for
third-party developers to write functions which take some formal
description of a set of values---examples include type annotations,
regular or context-free grammars, validation callbacks, and
JSONschema or GraphQL schemas---and return a strategy which generates
examples.

Several projects have chosen to fuzz web APIs with Hypothesis because of
the ease and effectiveness of sophisticated input generation, including
\texttt{swagger-fuzzer} \cite{swagger-fuzzer}
\texttt{swagger-conformance} \cite{swagger-conformance},
Yelp's \cite{REST-ler}-inspired \texttt{fuzz-lightyear} \cite{fuzz-lightyear}
for insecure direct object reference vulnerabilities,
and finally Schemathesis itself.

\subsection{Web API standards}

The REST architecture was described in 2000 \cite{REST}, with the
Swagger/OpenAPI schema format subsequently invented in 2011.
While OpenAPI is far from the \emph{only} specification format for REST
APIs, it is by far the most common.  Thanks to the shared architecture
and semantics, schemas in other formats\footnote{such as RAML, WSDL, or API Blueprint}
can be quickly and often automatically translated into the OpenAPI format.

The GraphQL architecture and schema format were designed together at Facebook
from 2012, and released in 2015 \cite{GraphQL}. Motivated by performance
considerations, GraphQL uses a fundamentally different data and request model
to RESTful APIs, but the same property-based testing techniques are applicable
to each.

While system state and highly structured inputs make fuzzing challenging,
OpenAPI or GraphQL schemas make it practical to exploit this structure.
Combined with HTTP's clear client/service interface and common properties
arising from the standardised semantics, schema-based fuzzing has an
enduring attraction.

\subsection{Standards imply semantic properties}
\label{sec:semantic-properties}

In addition to input structure, schemas constrain application semantics.
Wherever standards documents impose requirements on implementations,
semantics-aware fuzzers can treat ``requirements not violated'' as a
testable property---including \texttt{RFC 7231}'s universal
and relatively simple constraints on HTTP services, such as:
\begin{itemize}
\item
  \texttt{200\ OK} responses must have a non-empty body
\item
  \texttt{204\ No\ Content} and \texttt{205\ Reset\ Content} responses
  must have an empty body
\item
  \texttt{302\ Found} responses to a \texttt{POST} request must allow
  the subsequent response to use either \texttt{POST} or \texttt{GET}
  methods
\item
  \texttt{405\ Method\ Not\ Allowed} responses must have an
  \texttt{Allow} header listing supported methods
\item
  \texttt{500\ Internal\ Server\ Error} responses are always errors
\item
  \texttt{GET} fails after successful \texttt{DELETE}
  (use-after-free rule) \cite{CheckingRESTsecurity}
\item
  \texttt{GET} fails after unsuccessful \texttt{POST}
  (resource-leak rule) \cite{CheckingRESTsecurity}
\end{itemize}
In addition to input structure, API schemas document the expected
content and structure of API responses and the relationship between
endpoints. OpenAPI schemas extend HTTP semantics to include further
schema-specific testable properties, including:
\begin{itemize}
\item
  Response has undeclared status code
\item
  Response has undeclared content-type
\item
  Response body matches the schema
\item
  Response has wrong headers or missing required headers\footnote{
    Schemathesis seems to be the only fuzzer to check
    if required headers are missing.}
\item
  Non-conforming requests are rejected (negative testing) \cite{RESTTESTGEN}
\item
  No information leaks from unauthorised requests \cite{CheckingRESTsecurity,fuzz-lightyear}
\end{itemize}
Information disclosure vulnerabilities such as insecure direct object
references can be detected by making two sequences of
requests. The ``victim'' sequence creates and then retrieves a private
resource; then the ``attacker'' creates their own resource but attempts
to retrieve the \textit{victim} resource\footnote{
  A property relating multiple
  input-output pairs is known as a \emph{metamorphic relation}.
  \cite{MetamorphicSecurityTesting} propose twenty-two such relations
  relevant to web application security, which are often easier to check
  than one-shot properties.  See also \S\ref{sec:req-sequences}.
}. While in principle this could be done purely
on the basis of OpenAPI's \texttt{security} and \texttt{securitySchemes}
keywords, it is typically customised on a per-service basis.

Finally, we might wish to check for performance problems which could be
exploited to mount a denial-of-service attack. These checks are
typically disabled by default, because failure consists of exceeding a
subjective and configurable threshold rather than a binary
success/failure state:
\begin{itemize}
\item
  Slow responses, i.e. time to first (or last) byte of response
\item
  Request amplification measured in number of requests, or
  total response size, or amplification ratio
\end{itemize}
Hypothesis' support for \emph{targeted} property-based testing
\cite{TargetedPBT,AutomatingTargetedPBT} is particularly valuable for threshold tests
\cite{FalsifyYourSoftware}, and targets may be reported either by
Schemathesis derived fuzzers, or in user-supplied hooks. Anecdotally, we
have found that multi-objective optimisation---e.g., targeting response
size \emph{and} amplification ratio---is remarkably effective.

\subsection{Prior art in schema-based web API fuzzing}

Property-based testing provides the most common framing for
schema-based web API fuzzing, with early work such as \cite{QuickcheckWSDL,Jsongen}
focused on derived input generators (structure) and others
\cite{JsongenStateful,PropErWSDL,RESTest},
additionally deriving test oracles (semantics).

QuickREST \cite{QuickREST} aims to free up human effort by automating
API exploration with \texttt{Clojure.spec}.  Their tool tests for response
conformance and known HTTP status codes, with further properties specifiable
by the user.

\textsc{RestTestGen} \cite{RESTTESTGEN} proposes heuristics to statically recover an
`operation dependency graph' which describes the relationship between
response data and subsequent request to other endpoints - and enables far
more efficient testing with sequences of requests.
They also comment on the importance of testing error handling, i.e. negative
testing, using the property that nonconforming requests should receive
HTTP 4xx instead of 2xx responses.

RESTler \cite{REST-ler,CheckingRESTsecurity} is a  security testing tool in the tradition
of greybox fuzzing, which infers dependencies between request types, generates sequences
of requests satisfying those dependencies, and learns to predict sequence
validity based on the responses to these test sequences.  Their evaluation
finds that all these features are required for effective testing of web APIs.

EvoMaster \cite{EvoMaster} frames the problem as one of Evosuite-style
\cite{EvoSuite} \textit{test-case generation}, using JVM instrumentation to
evolve a high-coverage set of test cases.  A blackbox mode supports
non-JVM services, albeit with reduced performance.
\cite{EdDouibi2018ModelBased}  also generates JUnit test suites, based
on a model of schema and test semantics rather than execution feedback.

We are unaware of prior work on GraphQL schema-based testing
beyond Karlsson et al \cite{karlsson2020automatic}, who
demonstrate a proof-of-concept property-based testing tool
and a method for evaluating the schema coverage achieved.

To summarise, an ideal feature set might include:
\begin{itemize}
\item
  Deriving input generators and test oracles from API schemas,
  for both valid and invalid actions and data
\item
  Some kind of feedback to enable search-based testing, ideally
  available in the blackbox HTTP-only setting
\item
  A way to make sequences of requests, and exploit the data in
  responses to make further semantically-meaningful requests
\item
  A way to learn or discover relationships between endpoints
  for more efficient sequence-of-requests testing\footnote{
  e.g. \cite{RESTTESTGEN}'s operation dependency graph.
  Schemathesis relies on OpenAPI 3.0 ``links'' for this purpose,
  which are often omitted from schemas despite their value to tools.}
\end{itemize}

\section{Schemathesis}

Schemathesis is a tool for automated blackbox or whitebox randomised
testing of web APIs, deriving structure- and semantics-aware fuzzers from
OpenAPI or GraphQL schemas.

Derived fuzzers use Hypothesis' \cite{MacIver2019}
mature and sophisticated toolkit for creating test inputs---including hybrid
random generation, feedback-guided structured mutation, and explicit
examples---plus a variety of test functions and oracles for both individual
endpoints and sequences of requests to multiple endpoints.  Both data-generation
and test oracles can easily be customised by the end user.

Schemathesis can be used via a command-line interface or a Python API, in
either case to fuzz services written in any language via HTTP.
If the service under test is \emph{also} written in Python,
Schemathesis can communicate in-process using the WSGI and ASGI\footnote{
\href{https://wsgi.readthedocs.io/}{Web Server Gateway Interface}
and \href{https://asgi.readthedocs.io/}{Asynchronous Server Gateway Interface}}
conventions rather than over the network. This is often considerably
faster, and supports coverage-guided fuzzing with tools such as HypoFuzz\footnote{
\url{https://hypofuzz.com/}}, or Atheris\footnote{
\url{https://github.com/google/atheris}} via Hypothesis'
\texttt{fuzz\_one\_input} interface.

With thousands of downloads every week, Schemathesis is a thriving open-source
project.  It has already been widely adopted, integrated into Microsoft's REST API
Fuzz Testing project\footnote{\url{https://github.com/microsoft/rest-api-fuzz-testing}},
training from Red Hat\footnote{\url{https://appdev.consulting.redhat.com/tracks/contract-first/}},
and is the basis for the IBM Service Validator
\cite{ibm-service-validator}.

\subsection{Single requests or sequences?}
\label{sec:req-sequences}

Internally, Schemathesis distinguishes single-request tests from those
which make a sequence of requests to multiple endpoints, using the data
from past responses.  In 2020, we converted the latter from special-purpose
logic to a thin wrapper around Hypothesis' \texttt{RuleBasedStateMachine},
a generic system defined in terms of transition rules between states\footnote{
Similar to \cite{JsongenStateful}'s use of QuiviQ state machines}.
Comparing these implementations gives us some basis to discuss the
advantages of property-based testing.

The new state-machine tests tend to report \emph{fewer} bugs per run, because
they test the whole system rather than individual endpoints, but are
correspondingly faster to run. The difference is fundamentally a
property-based testing vs. a fuzzing workflow: the new style is designed
for interactive use in a run-fix-rerun cycle, rather than long-running
testing campaigns.

We believe that the heuristics built into Hypothesis, including swarm
testing \cite{SwarmTesting}, make a substantial contribution to
Schemathesis' performance. While such
techniques can be added to standalone tools, our experience is that the
implementation effort is best shared and tuned in dedicated
property-based testing libraries which benefit from synergies between
techniques.
\footnote{
For example, Hypothesis is to our knowledge the only tool to support
both swarm testing and targeted property-based testing
\cite{SwarmTesting,TargetedPBT,AutomatingTargetedPBT}.
}

\begin{table}[tb]
  \centering
  \begin{tabular}{lrrrrrrrrrr}
    
        & \rotatebox{90}{cccatalog} & \rotatebox{90}{covid19\_japan} & \rotatebox{90}{disease\_sh} & \rotatebox{90}{jupyter\_server} & \rotatebox{90}{jupyterhub} & \rotatebox{90}{open\_fec} & \rotatebox{90}{request\_baskets} & \rotatebox{90}{restler\_demo} & \rotatebox{90}{worklog} & \rotatebox{90}{age\_of\_empires} \\
    
        State-machine & \cellcolor[rgb]{0.935747,0.423831,0.13544} \color{white}3 & \cellcolor[rgb]{0.988362,0.998364,0.644924} 0 & \cellcolor[rgb]{0.988362,0.998364,0.644924} 0 & \cellcolor[rgb]{0.769556,0.236077,0.307485} \color{white}8 & \cellcolor[rgb]{0.935747,0.423831,0.13544} \color{white}3 & \cellcolor[rgb]{0.176493,0.041402,0.348111} \color{white}78 & \cellcolor[rgb]{0.988362,0.998364,0.644924} 0 & \cellcolor[rgb]{0.902003,0.364492,0.184116} \color{white}4 & \cellcolor[rgb]{0.987714,0.682807,0.072489} 1 & \cellcolor[rgb]{0.987714,0.682807,0.072489} 1 \\
    Hand-written &  & \cellcolor[rgb]{0.988362,0.998364,0.644924} 0 & \cellcolor[rgb]{0.988362,0.998364,0.644924} 0 & \cellcolor[rgb]{0.747127,0.222378,0.322856} \color{white}9 & \cellcolor[rgb]{0.987714,0.682807,0.072489} 1 & \cellcolor[rgb]{0.169575,0.042489,0.340874} \color{white}80 & \cellcolor[rgb]{0.988362,0.998364,0.644924} 0 & \cellcolor[rgb]{0.902003,0.364492,0.184116} \color{white}4 & \cellcolor[rgb]{0.987714,0.682807,0.072489} 1 & \cellcolor[rgb]{0.987714,0.682807,0.072489} 1 \\
    
    \bottomrule
  \end{tabular}
\caption{Switching to Hypothesis' state-machines made Schemathesis considerably 
faster and improved the quality of reported failures. In this evaluation on 
schemas we augumented with OpenAPI `links', state-machine tests detected a 
similar number of defects the old hand-written approach.
See Section \ref{sec:defect-detection} for details.
}
\label{table:testing-stateful}
\end{table}

\subsection{Hypothesis-Jsonschema}

Schemathesis converts schemas into data generators using the
\texttt{hypothesis-jsonschema}\footnote{\url{https://pypi.org/project/hypothesis-jsonschema/}}
library, whose \texttt{from\_schema()} function takes an arbitrary
JSON Schema\footnote{\url{https://json-schema.org/}} and returns a
Hypothesis strategy to generate valid instances. JSONschema is something
of a \textit{lingua franca} for web related schemas; Swagger and OpenAPI
use it directly, while others are easy to convert into JSONschemas.
Simple schemas admit simple translations:
\begin{verbatim}
    {"type": "integer", "minimum": 0, "maximum": 10}
    -> st.integers(min_value=0, max_value=10)
\end{verbatim}
while others, especially if they involve the \texttt{oneOf} or
\texttt{allOf}\footnote{
    \texttt{anyOf} is trivially satisfied by taking the union of the
    generators for sub-schemas.

    \texttt{oneOf} admits a quadratically-large translation to
    \texttt{allOf}, \texttt{not}, and \texttt{anyOf}---e.g. \texttt{oneOf: [a, b, ...]}
    $\rightarrow$ \texttt{anyOf: [\{a, \{not: \{anyOf: [b, ...]\}\}\}, ...]}---but
    linear-size rewrites and rejection sampling are usually much faster in practice.
}
combinators, defy easy or efficient translation.
We therefore `canonicalise' schemas by defining a suite of rewrite rules
which preserve schema semantics while reducing the need for rejection
sampling, and iterate them to a fixpoint.  Consider for example:
\begin{verbatim}
    {"type": "object", "allOf": [
         {"additionalProperties": false},
         {"properties": {"a": {"type": "string"}}},
     ]}
\end{verbatim}
containing intersecting constraints - the value must be an object, must not contain
any items if it is an object, and if the value is an object with key \texttt{"a"},
the corresponding value must be a string.
Our rewrite rules combine these constraints into a single schema:
\begin{verbatim}
    {"additionalProperties": False,
     "properties": {"a": False},
     "type": "object",
     "maxProperties": 1}
\end{verbatim}
and further simplify that schema into a minimal form:
\begin{verbatim}
    {'type': 'object', 'maxProperties': 0}
\end{verbatim}

Such nonlocal constraints are common in real-world schemas even before accounting
for widespread use of the \texttt{\$ref} keyword. We therefore inline
non-recursive references, merge overlapping subschemas, and canonicalise
the results before converting the schema to a Hypothesis strategy, which makes
\texttt{hypothesis-jsonschema} considerably faster than na\"ive translators such
as \texttt{Jsongen} \cite{Jsongen} with an otherwise similar design.

As well as generating valid inputs with no relation to ordinary
production traffic, this logic gives us an elegant way to synthesise
\emph{subtly invalid examples} for `negative testing' to check that
nonconforming requests are rejected by the API: create a set of
variant schemas, e.g. ``an invalid instance of a valid type'', and
then use each of the variants to generate input:
\begin{verbatim}
    schema = ...
    gen_invalid = from_schema(
        {"type": schema["type"], "not": schema}
    )
\end{verbatim}
This generator will be as efficient as anything we could write by hand\footnote{
    we designed comprehensive rewrite rules based on our knowledge of the spec,
    and update or expand them when inefficiencies are reported, e.g. by Reviewer \#2 (really!)
}, guaranteeing that instances are invalid with a minimum of rejection sampling.
By contrast, RESTTESTGEN \cite{RESTTESTGEN} generates data for negative testing
by applying relatively crude mutations to valid instances.

\subsection{Customising Schemathesis}

One of Schemathesis' strengths is ease of customisation via our
command-line interface or from Python.  The four main ways to
customise tests for a specific API are \emph{hooks}, \emph{checks},
\emph{serialisers}, and \emph{format strategies}.

\paragraph{Hooks} call user-defined functions to customise Schemathesis'
behaviour at different steps of the testing process:

\begin{itemize}

\item
  Hooks such as \texttt{before\_process\_path} allow you change the
  API schema for certain endpoints, working around incompatibilities or
  changing the data that will be generated
\item
  Hooks like \texttt{before\_generate\_query} allow you to replace
  Hypothesis strategies that are inferred from API schemas, for example
  by adding a filter to reject undesired test cases
\item
  Network request hooks allow you to send additional custom test cases,
  or to adjust generated data before sending it to the application under test
\item
  Custom targets, for e.g. performance tests as in \S\ref{sec:semantic-properties}
\end{itemize}

\paragraph{Checks}
are custom test oracles, which allow verification of user-defined properties
of responses received from the application under test.  Because checks are
decoupled from data generation, they can be run for both known-valid and
known-invalid test cases.

\paragraph{Serialisers}
Generated data must be serialised before transmission to to the application
under test.  Schemathesis supplies default serialisers for common media types
such as \texttt{application/json}, \texttt{multipart/form-data}, and
\texttt{text/plain}.  Custom serialisers can overide these defaults, or
add support for less common media types expected by the application.

\paragraph{Format strategies}
Many Open API schemas use custom \texttt{format} keywords to describe
the input data. For example, if the API under test consumes data that is
expected to contain a payment card number, this might be expressed as:
\begin{verbatim}
    {"type": "string", "format": "payment_card"}
\end{verbatim}
The JSONschema specification requires implementations to ignore unknown
\texttt{format} keys, so Schemathesis allows the user to supply a strategy
that will be used to generate values for this format:
\begin{verbatim}
    def luhn_ok(card_number: str) -> bool:
        """Validate check digit for the card number."""

    schemathesis.register_string_format(
        "payment_card",
        st.from_regex(r"^4[0-9]{15}$").filter(luhn_ok)
    )
\end{verbatim}

\subsection{The limits of specification support} \label{sssec:spec-support}

\texttt{hypothesis-jsonschema} correctly, and almost always efficiently, handles
every construct from draft-04 to draft-07 of the JSONschema specification
except for recursive references.
It's tested against every schema in the official JSONschema compatibility
test suite, the hundreds of real-world schemas from
\href{https://www.schemastore.org/json/}{schemastore.org}, and fuzzed with a
custom schema generator---to the point of finding bugs in Python's \texttt{jsonschema}
validator library and omissions from the standard-compliance test suite.\footnote{
    See \href{https://github.com/Julian/jsonschema/issues?q=author\%3AZac-HD+label\%3ABug}
    {Julian/jsonschema/issues?q=author\%3AZac-HD+label\%3ABug} \\
    and \href{https://github.com/json-schema-org/JSON-Schema-Test-Suite/pulls?q=author\%3AZac-HD}
    {json-schema-org/JSON-Schema-Test-Suite/pulls?q=author\%3AZac-HD}
}

Schemathesis supports almost everything in the OpenAPI spec, covering the
subset in common use and including user-defined extensions.
Of the more than three thousand schemas in the OpenAPI
directory, we successfully parse a higher proportion than are actually
valid---and can generate data for some endpoints of many invalid schemas.
Table \ref{table:OpenAPI-support} shows a detailed breakdown of the endpoints
we do \textit{not} support; aside from (some) recursive references, more
endpoints are \emph{invalid} than \emph{unsupported}.

\begin{table}
  \centering
  \begin{tabular}{lrc}
    \toprule
    Category & Number & Should work? \\
    \midrule
    Total OpenAPI directory endpoints & 66,925 & --- \\
    Schemathesis correctly handles & 65,195 & --- \\
    \midrule
    With unhandled recursive refs & 1,254 & yes \\
    With Python-incompatible regex & 170 & opt \\
    With too complex schemas & 26 & yes \\
    With un-inlined remote refs & 8 & yes \\
    \midrule
    With YAML parsing issues & 168 & no \\
    With invalid enums & 45 & no \\
    With path parameters including \texttt{/} & 33 & no \\
    With logically unsatisfiable schemas & 26 & no \\
    \bottomrule
  \end{tabular}
\caption{OpenAPI Directory endpoints, with detailed breakdown of those Schemathesis does not handle.}
\label{table:OpenAPI-support}
\end{table}

\begin{table*}[t]
  \centering
  \begin{tabular}{rl}

    \toprule
    Issue & Description \\
    \midrule

    \href{https://github.com/encode/django-rest-framework/issues/7134}{django-rest-framework\#7134} \href{https://github.com/encode/django-rest-framework/issues/7448}{and \#7448} &
    Crash on SQLite integer overflow due to validator order \\
    \href{https://github.com/Styria-Digital/django-rest-framework-jwt/issues/70}{django-rest-framework-jwt\#70} &
    Internal server error when token is invalid unicode \\
    \href{https://github.com/tfranzel/drf-spectacular/issues/186}{tfranzel/drf-spectacular\#186} &
    Generated schema is too permissive \\
    \href{https://github.com/redraw/satellite-passes-api/pull/2}{satellite-passes-api\#2} \href{https://github.com/redraw/satellite-passes-api/pull/4}{and \#4} &
    Internal server errors on missing attribute, missing cache key \\
    \href{https://github.com/tournesol-app/tournesol-backend/pull/17\#discussion_r669219677}{tournesol-backend\#17} &
    Checking HTTP status codes found ``a lot of bugs'' \\
    \href{https://github.com/OfficiumDivinum/OfficiumDivinum/issues/4}{OfficiumDivinum\#4} &
    Eleven failing tests with Schemathesis \\
    \href{https://github.com/python-restx/flask-restx/issues/303}{python-restx/flask-restx\#303} &
    Non-matching \texttt{X-} fields filter out all results instead of no results \\
    \href{https://github.com/http-rs/async-h1/issues/144}{http-rs/async-h1\#144} &
    Numerous client-side failures, causes unclear \\
    \href{https://github.com/tiangolo/fastapi/issues/240}{tiangolo/fastapi\#240} &
    Invalid schema (following wrong version of specification) \\
    \href{https://github.com/tiangolo/fastapi/issues/3790}{tiangolo/fastapi\#3790} &
    Nonconforming response when reporting validation errors \\
    \href{https://github.com/goadesign/goa/issues/2840}{goadesign/goa\#2840} &
    Schema is missing known HTTP status codes \\
    \href{https://github.com/Materials-Consortia/optimade-python-tools/issues/763}{optimade-python-tools\#763} &
    Schema is missing known HTTP status codes \\
    \href{https://github.com/marshmallow-code/apispec/issues/614}{marshmallow-code/apispec\#614} &
    Invalid schema attempting to bound datetime strings \\
    \href{https://github.com/jyveapp/django-action-framework/issues/14}{django-action-framework\#14} &
    Internal server error on unexpected URL fragment \\
    \href{https://github.com/jupyter-server/jupyter_server/issues/518}{jupyter-server\#518} &
    Schemathesis motivates and rewards comprehensive schemas \\

    \bottomrule

  \end{tabular}
  \caption{A sample of GitHub issues reporting independent use of Schemathesis.}
  \label{table:in-the-wild}
\end{table*}

Schemathesis currently handles recursive references by unrolling and inlining the
schema up to a reasonable depth.  This is sufficient for more than 98\% of endpoints
in the OpenAPI directory; but also the most common reason we fail to generate data.
We'd prefer to leverage property-based testing by expressing recursive schemas
directly with Hypothesis' \texttt{deferred()} generator, but modifying
\texttt{hypothesis-jsonschema}'s rewrite rules to be reference-aware is a tricky
engineering challenge---and unrolling works well enough that to date other features
have always taken priority.

We hypothesize that OpenAPI schemas found in the wild have a long tail
of rare features and complex structures, such that most non-trivial
schemas include at least one which breaks na\"ive fuzzers.

\section{Evaluation}

We experimentally evaluate Schemathesis'---and previous web API fuzzers---defect
detection, runtime, and consistency of reporting.
These experiments are restricted to containerised open-source services,
ensuring that they are representative, reproducible, and do not attack
live systems.

Since Schemathesis was open-sourced in August 2019, a range of
bug reports on GitHub have been attributed to Schemathesis.
A representative sample is shown in Table \ref{table:in-the-wild}
to contextualize our experiments.

Users evidently value detection of server errors (caused by conforming
inputs or not), as well as reporting of invalid, incomplete, or
overly-permissive schemas. Almost all of these results require the
fuzzer to understand schema semantics, in addition to the structure of
inputs and possible actions, to a higher degree than ``\texttt{HTTP\ 500}
means failure''.

\subsection{Experiment design}

\begin{table}[htb]
  \centering
  \begin{tabular}{llll}
    \toprule
    Name & Version & Language & Supported schemas \\
    \midrule
    Schemathesis & 3.9.0 & Python & Open API, GraphQL \\
    Restler & 7.1.0 & Python, F\# & Open API 2 / 3 \\
    Cats & 5.2.3 & Java & Open API 2 / 3 \\
    TnT-Fuzzer & 2.3.1 & Python & Open API 2 \\
    Got-Swag & 1.3.0 & JavaScript & Open API 2 \\
    APIFuzzer & b786c1b & Python & Open API 2 \\
    Fuzz-lightyear & 0.0.9 & Python & Open API 2 \\
    Swagger-conform & 0.2.5 & Python & Open API 2 \\
    Fuzzy Swagger & 0.1.11 & Python & Open API 2 \\
    Swagger fuzzer & 0.1.0 & Python & Open API 2 \\
    \bottomrule
  \end{tabular}
\caption{Evaluated schema-based web API fuzzers.}
\label{table:tools}
\end{table}

\begin{table*}[htb]
  \centering
  \begin{tabular}{lllcll}
    \toprule
    Service & Language & Framework & Endpoints & Schema type & Schema source \\
    \midrule

    \href{https://github.com/aalises/age-of-empires-II-api}{aalises/age-of-empires-II-api}
    & Python & Flask 1.1.2 & 8 & Open API 3.0.0 & Static \\
    \href{https://github.com/creativecommons/cccatalog-api}{creativecommons/cccatalog-api}
    & Python & Django 2.2.13 & 8 & Swagger 2.0 & Dynamic, drf-yasg 1.17.1 \\
    \href{https://github.com/ryo-ma/covid19-japan-web-api}{ryo-ma/covid19-japan-web-api}
    & Python & Flask 1.1.2 & 4 & Swagger 2.0 & Dynamic, flasgger 0.9.4 \\
    \href{https://github.com/disease-sh/api}{disease-sh/api}
    & JavaScript & Express 4.17.1 & 34 & Swagger 2.0 & Static \\
    \href{https://github.com/postmanlabs/httpbin}{postmanlabs/httpbin}
    & Python & Flask 1.0.2 & 73 & Swagger 2.0 & Dynamic, flasgger 0.9.0 \\
    \href{https://github.com/jupyter-server/jupyter\_server}{jupyter-server/jupyter\_server}
    & Python & Tornado 6.1.0 & 29 & Swagger 2.0 & Static \\
    \href{https://github.com/jupyterhub/jupyterhub}{jupyterhub/jupyterhub}
    & Python & Tornado 6.1.0 & 35 & Swagger 2.0 & Static \\
    \href{https://github.com/mailhog/MailHog}{mailhog/MailHog}
    & Go & Net/HTTP & 2 & Swagger 2.0 & Static \\
    \href{https://github.com/fecgov/openFEC}{fecgov/openFEC}
    & Python & Flask 1.1.1 & 85 & Swagger 2.0 & Dynamic, flask-apispec 0.7.0 \\
    \href{https://github.com/ajnisbet/opentopodata/}{ajnisbet/opentopodata/}
    & Python & Flask 1.1.2 & 2 & Open API 3.0.2 & Static \\
    \href{https://github.com/rtyler/otto}{rtyler/otto}
    & Rust & Tide 0.14.0 & 2 & Open API 3.0.3 & Static \\
    \href{https://github.com/fossasia/pslab-webapp}{fossasia/pslab-webapp}
    & Python & Flask 1.1.2 & 3 & Swagger 2.0 & Dynamic, flasgger 0.9.5 \\
    \href{https://github.com/pulp/pulpcore}{pulp/pulpcore}
    & Python & Django 2.2.17 & 67 & Open API 3.0.3 & Dynamic, drf-spectacular 0.11.0 \\
    \href{https://github.com/darklynx/request-baskets}{darklynx/request-baskets}
    & Go & Net/HTTP & 20 & Swagger 2.0 & Static \\
    \href{https://github.com/microsoft/restler-fuzzer}{microsoft/restler-fuzzer}
    & Python & Flask 1.1.2 & 6 & Swagger 2.0 & Static \\
    \href{https://github.com/IBM/worklog}{IBM/worklog}
    & Python & Flask 1.0.2 & 9 & Swagger 2.0 & Dynamic, flasgger 0.9.1 \\

    \bottomrule
  \end{tabular}
\caption{Tested web services, chosen to represent a variety of programming
languages, schema formats (OpenAPI and GraphQL), and sources (generated or hand-written)
across a realistic range of sizes, structures, and API complexity.}
\label{table:targets}
\end{table*}

We run three configurations of Schemathesis, and seven other fuzzing
tools in their default configurations
\footnote{mostly - following \cite{Fuzz2020}, we avoided fuzzing e.g.
Jupyter's \texttt{/shutdown} endpoint} (Table \ref{table:tools}),
on sixteen open-source web APIs (Table \ref{table:targets}),
for thirty runs each.

To our knowledge, this is the most comprehensive evaluation of web API
fuzzers to date. Our scripts make it easy to add further fuzzing tools
or targets in Docker containers, and will be maintained as a standard
benchmark suite for the community.

The full set of containers to reproduce our work, or use it in
evaluating future fuzzers, is available from
\href{https://github.com/schemathesis/web-api-fuzzing-project}
{github.com/schemathesis/web-api-fuzzing-project},
along with both raw and processed data.

For ease of analysis, we parse the 250GB of raw logs into a JSON
summary, and further reduce this dataset to report the duration, number
of events, and per-run reports of each unique defect.

Manual defect triage and deduplication is impractical for such a large
and extensible evaluation. Instead, where possible we monitor the
fuzzing process using Sentry\footnote{\url{https://sentry.io/}}, a
widely-used platform for error tracking and performance monitoring. This
gives us a cross-language notion of unique defects, i.e. internal server
errors deduplicated by code location---regardless of triggering endpoint
or what the fuzzer was attempting to check at the time.

We add semantic errors\footnote{unexpected status code, schema non-conformance,
information disclosure, etc.} to our defect count by counting each kind of bug
report parsed from saved logs only once per endpoint, regardless of variations or
how many times it was observed.  This matches our experience of users' tend
to group reports which are consistently either fixed or ignored.

This typically works well, with the notable exception
of TnT-Fuzzer---which reports more than a thousand \texttt{404\ Not\ Found}
responses \emph{for randomly-generated paths}:
\begin{verbatim}
    "result": {"type": "failure",
               "kind": "unexpected_status_code",
               "status_code": 404},
    "path": "/haugixqhzhjcgdwjqluktpxtzuerizamqmhqsu...
\end{verbatim}
We therefore count any number of ``unexpected 404'' reports as a single
unique defect per target, ignoring the path.

\subsection{Defect-detection experiment}
\label{sec:defect-detection}

\begin{table*}[ptbh]
  \centering
  \begin{tabular}{lrrrrrrrrrrrrrrrr}
    
    & \rotatebox{90}{cccatalog} & \rotatebox{90}{covid19\_japan} & \rotatebox{90}{disease\_sh} & \rotatebox{90}{httpbin} & \rotatebox{90}{jupyter\_server} & \rotatebox{90}{jupyterhub} & \rotatebox{90}{mailhog} & \rotatebox{90}{open\_fec} & \rotatebox{90}{pslab\_webapp} & \rotatebox{90}{request\_baskets} & \rotatebox{90}{restler\_demo} & \rotatebox{90}{worklog} & \rotatebox{90}{age\_of\_empires} & \rotatebox{90}{opentopodata} & \rotatebox{90}{otto\_parser} & \rotatebox{90}{pulpcore} \\
    \toprule
    OpenAPI version & 2 & 2 & 2 & 2 & 2 & 2 & 2 & 2 & 2 & 2 & 2 & 2 & 3 & 3 & 3 & 3 \\
    Endpoints & 8 & 4 & 34 & 73 & 29 & 35 & 2 & 85 & 3 & 20 & 6 & 9 & 8 & 2 & 2 & 67 \\

    \toprule
    
    api\_fuzzer &  & \cellcolor[rgb]{0.988362,0.998364,0.644924} 0 & \cellcolor[rgb]{0.988362,0.998364,0.644924} 0 & \cellcolor[rgb]{0.391453,0.080927,0.433109} \color{white}36 &  &  & \cellcolor[rgb]{0.988362,0.998364,0.644924} 0 & \cellcolor[rgb]{0.590734,0.152563,0.40029} \color{white}17 & \cellcolor[rgb]{0.970919,0.522853,0.058367} 2 & \cellcolor[rgb]{0.603139,0.157151,0.395891} \color{white}16 &  & \cellcolor[rgb]{0.832299,0.283913,0.257383} \color{white}6 & \cellcolor[rgb]{0.902003,0.364492,0.184116} \color{white}4 & \cellcolor[rgb]{0.988362,0.998364,0.644924} 0 &  & \cellcolor[rgb]{0.700576,0.197851,0.351113} \color{white}11 \\
    cats &  & \cellcolor[rgb]{0.988362,0.998364,0.644924} 0 & \cellcolor[rgb]{0.988362,0.998364,0.644924} 0 &  & \cellcolor[rgb]{0.865006,0.316822,0.226055} \color{white}5 &  & \cellcolor[rgb]{0.988362,0.998364,0.644924} 0 & \cellcolor[rgb]{0.718264,0.206636,0.340931} \color{white}10 & \cellcolor[rgb]{0.970919,0.522853,0.058367} 2 & \cellcolor[rgb]{0.988362,0.998364,0.644924} 0 &  &  &  & \cellcolor[rgb]{0.988362,0.998364,0.644924} 0 &  & \cellcolor[rgb]{0.832299,0.283913,0.257383} \color{white}6 \\
    fuzz\_lightyear &  &  &  &  &  &  &  & \cellcolor[rgb]{0.970919,0.522853,0.058367} 2 & \cellcolor[rgb]{0.988362,0.998364,0.644924} 0 &  &  &  &  &  &  &  \\
    got\_swag & \cellcolor[rgb]{0.935747,0.423831,0.13544} \color{white}3 &  &  &  &  &  &  & \cellcolor[rgb]{0.422549,0.092501,0.432714} \color{white}32 &  &  & \cellcolor[rgb]{0.988362,0.998364,0.644924} 0 &  & \cellcolor[rgb]{0.988362,0.998364,0.644924} 0 &  &  & \cellcolor[rgb]{0.658463,0.178962,0.372748} \color{white}13 \\
    restler & \cellcolor[rgb]{0.970919,0.522853,0.058367} 2 & \cellcolor[rgb]{0.988362,0.998364,0.644924} 0 &  &  & \cellcolor[rgb]{0.987714,0.682807,0.072489} 1 &  & \cellcolor[rgb]{0.988362,0.998364,0.644924} 0 & \cellcolor[rgb]{0.935747,0.423831,0.13544} \color{white}3 & \cellcolor[rgb]{0.970919,0.522853,0.058367} 2 & \cellcolor[rgb]{0.988362,0.998364,0.644924} 0 & \cellcolor[rgb]{0.988362,0.998364,0.644924} 0 & \cellcolor[rgb]{0.988362,0.998364,0.644924} 0 & \cellcolor[rgb]{0.988362,0.998364,0.644924} 0 & \cellcolor[rgb]{0.988362,0.998364,0.644924} 0 &  &  \\
    schemathesis:AllChecks & \cellcolor[rgb]{0.658463,0.178962,0.372748} \color{white}13 & \cellcolor[rgb]{0.988362,0.998364,0.644924} 0 & \cellcolor[rgb]{0.700576,0.197851,0.351113} \color{white}11 & \cellcolor[rgb]{0.001462,0.000466,0.013866} \color{white}316 & \cellcolor[rgb]{0.422549,0.092501,0.432714} \color{white}32 & \cellcolor[rgb]{0.676638,0.186807,0.363849} \color{white}12 & \cellcolor[rgb]{0.987714,0.682807,0.072489} 1 & \cellcolor[rgb]{0.004547,0.003392,0.030909} \color{white}184 & \cellcolor[rgb]{0.801871,0.258674,0.283099} \color{white}7 & \cellcolor[rgb]{0.372768,0.073915,0.4324} \color{white}39 & \cellcolor[rgb]{0.640135,0.171438,0.381065} \color{white}14 & \cellcolor[rgb]{0.935747,0.423831,0.13544} \color{white}3 & \cellcolor[rgb]{0.801871,0.258674,0.283099} \color{white}7 & \cellcolor[rgb]{0.987714,0.682807,0.072489} 1 & \cellcolor[rgb]{0.988362,0.998364,0.644924} 0 & \cellcolor[rgb]{0.497257,0.119379,0.424488} \color{white}24 \\
    schemathesis:Default & \cellcolor[rgb]{0.935747,0.423831,0.13544} \color{white}3 & \cellcolor[rgb]{0.988362,0.998364,0.644924} 0 & \cellcolor[rgb]{0.988362,0.998364,0.644924} 0 & \cellcolor[rgb]{0.231538,0.036405,0.3924} \color{white}66 & \cellcolor[rgb]{0.718264,0.206636,0.340931} \color{white}10 & \cellcolor[rgb]{0.987714,0.682807,0.072489} 1 & \cellcolor[rgb]{0.988362,0.998364,0.644924} 0 & \cellcolor[rgb]{0.169575,0.042489,0.340874} \color{white}80 & \cellcolor[rgb]{0.970919,0.522853,0.058367} 2 & \cellcolor[rgb]{0.988362,0.998364,0.644924} 0 & \cellcolor[rgb]{0.902003,0.364492,0.184116} \color{white}4 & \cellcolor[rgb]{0.987714,0.682807,0.072489} 1 & \cellcolor[rgb]{0.987714,0.682807,0.072489} 1 & \cellcolor[rgb]{0.988362,0.998364,0.644924} 0 & \cellcolor[rgb]{0.988362,0.998364,0.644924} 0 & \cellcolor[rgb]{0.865006,0.316822,0.226055} \color{white}5 \\
    schemathesis:Negative & \cellcolor[rgb]{0.935747,0.423831,0.13544} \color{white}3 & \cellcolor[rgb]{0.988362,0.998364,0.644924} 0 & \cellcolor[rgb]{0.988362,0.998364,0.644924} 0 & \cellcolor[rgb]{0.658463,0.178962,0.372748} \color{white}13 & \cellcolor[rgb]{0.769556,0.236077,0.307485} \color{white}8 & \cellcolor[rgb]{0.366529,0.071579,0.431994} \color{white}40 & \cellcolor[rgb]{0.988362,0.998364,0.644924} 0 & \cellcolor[rgb]{0.197297,0.0384,0.367535} \color{white}73 & \cellcolor[rgb]{0.970919,0.522853,0.058367} 2 & \cellcolor[rgb]{0.988362,0.998364,0.644924} 0 & \cellcolor[rgb]{0.987714,0.682807,0.072489} 1 & \cellcolor[rgb]{0.935747,0.423831,0.13544} \color{white}3 & \cellcolor[rgb]{0.987714,0.682807,0.072489} 1 & \cellcolor[rgb]{0.988362,0.998364,0.644924} 0 & \cellcolor[rgb]{0.988362,0.998364,0.644924} 0 & \cellcolor[rgb]{0.747127,0.222378,0.322856} \color{white}9 \\
    swagger\_fuzzer &  &  &  &  &  &  &  & \cellcolor[rgb]{0.603139,0.157151,0.395891} \color{white}16 &  &  &  &  &  &  &  &  \\
    tnt\_fuzzer &  &  &  &  &  &  &  &  &  &  &  &  &  &  &  & \cellcolor[rgb]{0.718264,0.206636,0.340931} \color{white}10 \\
    
    \midrule
    \multicolumn{17}{l}{
        \parbox{0.95\linewidth}{
            \caption{Total number of unique defects across all 30 runs, regardless of triggering endpoint, using Sentry to determine `ground truth' for internal server errors and custom log parsing to group property violations.}
            \vspace{-0.8em}
            \label{table:uniq_failures}
    }} \\

    \toprule
    
    api\_fuzzer &  & \cellcolor[rgb]{0.988362,0.998364,0.644924} 0 & \cellcolor[rgb]{0.988362,0.998364,0.644924} 0 & \cellcolor[rgb]{0.658463,0.178962,0.372748} \color{white}13 &  &  & \cellcolor[rgb]{0.988362,0.998364,0.644924} 0 & \cellcolor[rgb]{0.590734,0.152563,0.40029} \color{white}17 & \cellcolor[rgb]{0.970919,0.522853,0.058367} 2 & \cellcolor[rgb]{0.988362,0.998364,0.644924} 0 &  & \cellcolor[rgb]{0.987714,0.682807,0.072489} 1 & \cellcolor[rgb]{0.988362,0.998364,0.644924} 0 & \cellcolor[rgb]{0.988362,0.998364,0.644924} 0 &  & \cellcolor[rgb]{0.988362,0.998364,0.644924} 0 \\
    cats &  & \cellcolor[rgb]{0.988362,0.998364,0.644924} 0 & \cellcolor[rgb]{0.988362,0.998364,0.644924} 0 &  & \cellcolor[rgb]{0.970919,0.522853,0.058367} 2 &  & \cellcolor[rgb]{0.988362,0.998364,0.644924} 0 & \cellcolor[rgb]{0.718264,0.206636,0.340931} \color{white}10 & \cellcolor[rgb]{0.970919,0.522853,0.058367} 2 & \cellcolor[rgb]{0.988362,0.998364,0.644924} 0 &  &  &  & \cellcolor[rgb]{0.988362,0.998364,0.644924} 0 &  & \cellcolor[rgb]{0.970919,0.522853,0.058367} 2 \\
    fuzz\_lightyear &  &  &  &  &  &  &  & \cellcolor[rgb]{0.987714,0.682807,0.072489} 1 & \cellcolor[rgb]{0.988362,0.998364,0.644924} 0 &  &  &  &  &  &  &  \\
    got\_swag & \cellcolor[rgb]{0.970919,0.522853,0.058367} 2 &  &  &  &  &  &  & \cellcolor[rgb]{0.865006,0.316822,0.226055} \color{white}5 &  &  & \cellcolor[rgb]{0.988362,0.998364,0.644924} 0 &  & \cellcolor[rgb]{0.988362,0.998364,0.644924} 0 &  &  & \cellcolor[rgb]{0.988362,0.998364,0.644924} 0 \\
    restler & \cellcolor[rgb]{0.970919,0.522853,0.058367} 2 & \cellcolor[rgb]{0.988362,0.998364,0.644924} 0 &  &  & \cellcolor[rgb]{0.987714,0.682807,0.072489} 1 &  & \cellcolor[rgb]{0.988362,0.998364,0.644924} 0 & \cellcolor[rgb]{0.935747,0.423831,0.13544} \color{white}3 & \cellcolor[rgb]{0.970919,0.522853,0.058367} 2 & \cellcolor[rgb]{0.988362,0.998364,0.644924} 0 & \cellcolor[rgb]{0.988362,0.998364,0.644924} 0 & \cellcolor[rgb]{0.988362,0.998364,0.644924} 0 & \cellcolor[rgb]{0.988362,0.998364,0.644924} 0 & \cellcolor[rgb]{0.988362,0.998364,0.644924} 0 &  &  \\
    schemathesis:AllChecks & \cellcolor[rgb]{0.832299,0.283913,0.257383} \color{white}6 & \cellcolor[rgb]{0.988362,0.998364,0.644924} 0 & \cellcolor[rgb]{0.988362,0.998364,0.644924} 0 & \cellcolor[rgb]{0.621685,0.164184,0.388781} \color{white}15 & \cellcolor[rgb]{0.621685,0.164184,0.388781} \color{white}15 & \cellcolor[rgb]{0.970919,0.522853,0.058367} 2 & \cellcolor[rgb]{0.988362,0.998364,0.644924} 0 & \cellcolor[rgb]{0.176493,0.041402,0.348111} \color{white}78 & \cellcolor[rgb]{0.902003,0.364492,0.184116} \color{white}4 & \cellcolor[rgb]{0.988362,0.998364,0.644924} 0 & \cellcolor[rgb]{0.832299,0.283913,0.257383} \color{white}6 & \cellcolor[rgb]{0.970919,0.522853,0.058367} 2 & \cellcolor[rgb]{0.988362,0.998364,0.644924} 0 & \cellcolor[rgb]{0.988362,0.998364,0.644924} 0 & \cellcolor[rgb]{0.988362,0.998364,0.644924} 0 & \cellcolor[rgb]{0.769556,0.236077,0.307485} \color{white}8 \\
    schemathesis:Default & \cellcolor[rgb]{0.935747,0.423831,0.13544} \color{white}3 & \cellcolor[rgb]{0.988362,0.998364,0.644924} 0 & \cellcolor[rgb]{0.988362,0.998364,0.644924} 0 & \cellcolor[rgb]{0.718264,0.206636,0.340931} \color{white}10 & \cellcolor[rgb]{0.747127,0.222378,0.322856} \color{white}9 & \cellcolor[rgb]{0.987714,0.682807,0.072489} 1 & \cellcolor[rgb]{0.988362,0.998364,0.644924} 0 & \cellcolor[rgb]{0.169575,0.042489,0.340874} \color{white}80 & \cellcolor[rgb]{0.970919,0.522853,0.058367} 2 & \cellcolor[rgb]{0.988362,0.998364,0.644924} 0 & \cellcolor[rgb]{0.902003,0.364492,0.184116} \color{white}4 & \cellcolor[rgb]{0.987714,0.682807,0.072489} 1 & \cellcolor[rgb]{0.988362,0.998364,0.644924} 0 & \cellcolor[rgb]{0.988362,0.998364,0.644924} 0 & \cellcolor[rgb]{0.988362,0.998364,0.644924} 0 & \cellcolor[rgb]{0.865006,0.316822,0.226055} \color{white}5 \\
    schemathesis:Negative & \cellcolor[rgb]{0.935747,0.423831,0.13544} \color{white}3 & \cellcolor[rgb]{0.988362,0.998364,0.644924} 0 & \cellcolor[rgb]{0.988362,0.998364,0.644924} 0 & \cellcolor[rgb]{0.700576,0.197851,0.351113} \color{white}11 & \cellcolor[rgb]{0.801871,0.258674,0.283099} \color{white}7 & \cellcolor[rgb]{0.832299,0.283913,0.257383} \color{white}6 & \cellcolor[rgb]{0.988362,0.998364,0.644924} 0 & \cellcolor[rgb]{0.197297,0.0384,0.367535} \color{white}73 & \cellcolor[rgb]{0.970919,0.522853,0.058367} 2 & \cellcolor[rgb]{0.988362,0.998364,0.644924} 0 & \cellcolor[rgb]{0.987714,0.682807,0.072489} 1 & \cellcolor[rgb]{0.935747,0.423831,0.13544} \color{white}3 & \cellcolor[rgb]{0.988362,0.998364,0.644924} 0 & \cellcolor[rgb]{0.988362,0.998364,0.644924} 0 & \cellcolor[rgb]{0.988362,0.998364,0.644924} 0 & \cellcolor[rgb]{0.747127,0.222378,0.322856} \color{white}9 \\
    swagger\_fuzzer &  &  &  &  &  &  &  & \cellcolor[rgb]{0.988362,0.998364,0.644924} 0 &  &  &  &  &  &  &  &  \\
    tnt\_fuzzer &  &  &  &  &  &  &  &  &  &  &  &  &  &  &  & \cellcolor[rgb]{0.988362,0.998364,0.644924} 0 \\
    
    \midrule
    \multicolumn{17}{l}{
        \parbox{0.95\linewidth}{
            \caption{Total unique \texttt{HTTP 500} server errors, for comparison between tools which check different semantic properties.}
            \vspace{-0.8em}
            \label{table:uniq_500s}
    }} \\

    \toprule
    
    api\_fuzzer &  & \cellcolor[rgb]{0.935904,0.89857,0.108131} 3 & \cellcolor[rgb]{0.237441,0.305202,0.541921} \color{white}23:32 & \cellcolor[rgb]{0.204903,0.375746,0.553533} \color{white}13:25 &  &  & \cellcolor[rgb]{0.964894,0.902323,0.123941} 2 & \cellcolor[rgb]{0.1941,0.399323,0.555565} \color{white}11:01 & \cellcolor[rgb]{0.319809,0.770914,0.411152} 23 & \cellcolor[rgb]{0.487026,0.823929,0.312321} 12 &  & \cellcolor[rgb]{0.252899,0.742211,0.448284} \color{white}31 & \cellcolor[rgb]{0.866013,0.889868,0.095953} 3 & \cellcolor[rgb]{0.82494,0.88472,0.106217} 4 &  & \cellcolor[rgb]{0.214298,0.355619,0.551184} \color{white}15:56 \\
    cats &  & \cellcolor[rgb]{0.762373,0.876424,0.137064} 5 & \cellcolor[rgb]{0.169646,0.456262,0.55803} \color{white}6:45 &  & \cellcolor[rgb]{0.515992,0.831158,0.294279} 11 &  & \cellcolor[rgb]{0.783315,0.879285,0.125405} 4 & \cellcolor[rgb]{0.119483,0.614817,0.537692} \color{white}1:42 & \cellcolor[rgb]{0.626579,0.854645,0.223353} 8 & \cellcolor[rgb]{0.762373,0.876424,0.137064} 5 &  &  &  & \cellcolor[rgb]{0.83527,0.886029,0.102646} 4 &  & \cellcolor[rgb]{0.119483,0.614817,0.537692} \color{white}1:42 \\
    fuzz\_lightyear &  &  &  &  &  &  &  & \cellcolor[rgb]{0.515992,0.831158,0.294279} 11 & \cellcolor[rgb]{0.993248,0.906157,0.143936} 1 &  &  &  &  &  &  &  \\
    got\_swag & \cellcolor[rgb]{0.993248,0.906157,0.143936} 1 &  &  &  &  &  &  & \cellcolor[rgb]{0.404001,0.800275,0.362552} 16 &  &  & \cellcolor[rgb]{0.993248,0.906157,0.143936} 1 &  & \cellcolor[rgb]{0.993248,0.906157,0.143936} 1 &  &  & \cellcolor[rgb]{0.993248,0.906157,0.143936} 1 \\
    restler & \cellcolor[rgb]{0.762373,0.876424,0.137064} 5 & \cellcolor[rgb]{0.906311,0.894855,0.098125} 3 &  &  & \cellcolor[rgb]{0.772852,0.877868,0.131109} 5 &  & \cellcolor[rgb]{0.916242,0.896091,0.100717} 3 & \cellcolor[rgb]{0.267004,0.004874,0.329415} \color{white}4:28:06 & \cellcolor[rgb]{0.876168,0.891125,0.09525} 3 & \cellcolor[rgb]{0.535621,0.835785,0.281908} 10 & \cellcolor[rgb]{0.804182,0.882046,0.114965} 4 & \cellcolor[rgb]{0.395174,0.797475,0.367757} 17 & \cellcolor[rgb]{0.886271,0.892374,0.095374} 3 & \cellcolor[rgb]{0.906311,0.894855,0.098125} 3 &  &  \\
    schemathesis:AllChecks & \cellcolor[rgb]{0.506271,0.828786,0.300362} 12 & \cellcolor[rgb]{0.9553,0.901065,0.118128} 2 & \cellcolor[rgb]{0.131172,0.555899,0.552459} \color{white}2:49 & \cellcolor[rgb]{0.243113,0.292092,0.538516} \color{white}25:51 & \cellcolor[rgb]{0.412913,0.803041,0.357269} 16 & \cellcolor[rgb]{0.360741,0.785964,0.387814} 20 & \cellcolor[rgb]{0.964894,0.902323,0.123941} 2 & \cellcolor[rgb]{0.190631,0.407061,0.556089} \color{white}10:32 & \cellcolor[rgb]{0.85581,0.888601,0.097452} 4 & \cellcolor[rgb]{0.647257,0.8584,0.209861} 7 & \cellcolor[rgb]{0.575563,0.844566,0.256415} 9 & \cellcolor[rgb]{0.220124,0.725509,0.466226} \color{white}37 & \cellcolor[rgb]{0.678489,0.863742,0.189503} 6 & \cellcolor[rgb]{0.162142,0.474838,0.55814} \color{white}5:48 & \cellcolor[rgb]{0.993248,0.906157,0.143936} 2 & \cellcolor[rgb]{0.147607,0.511733,0.557049} \color{white}4:16 \\
    schemathesis:Default & \cellcolor[rgb]{0.274149,0.751988,0.436601} \color{white}29 & \cellcolor[rgb]{0.983868,0.904867,0.136897} 2 & \cellcolor[rgb]{0.187231,0.414746,0.556547} \color{white}9:41 & \cellcolor[rgb]{0.267968,0.223549,0.512008} \color{white}42:29 & \cellcolor[rgb]{0.274149,0.751988,0.436601} \color{white}28 & \cellcolor[rgb]{0.137339,0.662252,0.515571} \color{white}1:06 & \cellcolor[rgb]{0.916242,0.896091,0.100717} 3 & \cellcolor[rgb]{0.179019,0.433756,0.55743} \color{white}8:18 & \cellcolor[rgb]{0.9553,0.901065,0.118128} 2 & \cellcolor[rgb]{0.220124,0.725509,0.466226} \color{white}37 & \cellcolor[rgb]{0.170948,0.694384,0.493803} \color{white}49 & \cellcolor[rgb]{0.232815,0.732247,0.459277} \color{white}35 & \cellcolor[rgb]{0.751884,0.874951,0.143228} 5 & \cellcolor[rgb]{0.916242,0.896091,0.100717} 3 & \cellcolor[rgb]{0.993248,0.906157,0.143936} 2 & \cellcolor[rgb]{0.182256,0.426184,0.55712} \color{white}8:52 \\
    schemathesis:Negative & \cellcolor[rgb]{0.162016,0.687316,0.499129} \color{white}53 & \cellcolor[rgb]{0.412913,0.803041,0.357269} 16 & \cellcolor[rgb]{0.151918,0.500685,0.557587} \color{white}4:36 & \cellcolor[rgb]{0.19586,0.395433,0.555276} \color{white}11:38 & \cellcolor[rgb]{0.170948,0.694384,0.493803} \color{white}50 & \cellcolor[rgb]{0.281924,0.089666,0.412415} \color{white}1:40:50 & \cellcolor[rgb]{0.430983,0.808473,0.346476} 15 & \cellcolor[rgb]{0.267004,0.004874,0.329415} \color{white}6:48:18 & \cellcolor[rgb]{0.935904,0.89857,0.108131} 3 & \cellcolor[rgb]{0.136408,0.541173,0.554483} \color{white}3:12 & \cellcolor[rgb]{0.239374,0.735588,0.455688} \color{white}33 & \cellcolor[rgb]{0.202219,0.715272,0.476084} \color{white}41 & \cellcolor[rgb]{0.636902,0.856542,0.21662} 8 & \cellcolor[rgb]{0.440137,0.811138,0.340967} 15 & \cellcolor[rgb]{0.678489,0.863742,0.189503} 6 & \cellcolor[rgb]{0.267968,0.223549,0.512008} \color{white}42:32 \\
    swagger\_fuzzer &  &  &  &  &  &  &  & \cellcolor[rgb]{0.709898,0.868751,0.169257} 6 &  &  &  &  &  &  &  &  \\
    tnt\_fuzzer &  &  &  &  &  &  &  &  &  &  &  &  &  &  &  & \cellcolor[rgb]{0.866013,0.889868,0.095953} 3 \\
    
    \midrule
    \multicolumn{17}{l}{
        \parbox{0.95\linewidth}{
            \caption{Mean runtime, showing a clear but noisy correlation to number of endpoints and detected defects.}
            \vspace{-0.8em}
            \label{table:durations}
    }} \\

    \toprule
    
    api\_fuzzer &  &  &  & \cellcolor[rgb]{0.28229,0.145912,0.46151} \color{white}34 &  &  &  & \cellcolor[rgb]{0.19586,0.395433,0.555276} \color{white}15 & \cellcolor[rgb]{0.267004,0.004874,0.329415} \color{white}130 & \cellcolor[rgb]{0.319809,0.770914,0.411152} 3 &  & \cellcolor[rgb]{0.277018,0.050344,0.375715} \color{white}44 & \cellcolor[rgb]{0.866013,0.889868,0.095953} 1 &  &  & \cellcolor[rgb]{0.267004,0.004874,0.329415} \color{white}1,276 \\
    cats &  &  &  &  & \cellcolor[rgb]{0.276022,0.044167,0.370164} \color{white}44 &  &  & \cellcolor[rgb]{0.192357,0.403199,0.555836} \color{white}15 & \cellcolor[rgb]{0.267004,0.004874,0.329415} \color{white}81 &  &  &  &  &  &  & \cellcolor[rgb]{0.449368,0.813768,0.335384} 2 \\
    fuzz\_lightyear &  &  &  &  &  &  &  & \cellcolor[rgb]{0.119512,0.607464,0.540218} \color{white}6 &  &  &  &  &  &  &  &  \\
    got\_swag & \cellcolor[rgb]{0.730889,0.871916,0.156029} 1 &  &  &  &  &  &  & \cellcolor[rgb]{0.993248,0.906157,0.143936} 1 &  &  &  &  &  &  &  & \cellcolor[rgb]{0.993248,0.906157,0.143936} 1 \\
    restler & \cellcolor[rgb]{0.616293,0.852709,0.230052} 2 &  &  &  & \cellcolor[rgb]{0.616293,0.852709,0.230052} 2 &  &  & \cellcolor[rgb]{0.616293,0.852709,0.230052} 2 & \cellcolor[rgb]{0.616293,0.852709,0.230052} 2 &  &  &  &  &  &  &  \\
    schemathesis:AllChecks & \cellcolor[rgb]{0.993248,0.906157,0.143936} 1 &  & \cellcolor[rgb]{0.120092,0.600104,0.54253} \color{white}7 & \cellcolor[rgb]{0.993248,0.906157,0.143936} 1 & \cellcolor[rgb]{0.983868,0.904867,0.136897} 1 & \cellcolor[rgb]{0.993248,0.906157,0.143936} 1 & \cellcolor[rgb]{0.993248,0.906157,0.143936} 1 & \cellcolor[rgb]{0.935904,0.89857,0.108131} 1 & \cellcolor[rgb]{0.993248,0.906157,0.143936} 1 & \cellcolor[rgb]{0.993248,0.906157,0.143936} 1 & \cellcolor[rgb]{0.9553,0.901065,0.118128} 1 & \cellcolor[rgb]{0.993248,0.906157,0.143936} 1 & \cellcolor[rgb]{0.993248,0.906157,0.143936} 1 & \cellcolor[rgb]{0.993248,0.906157,0.143936} 1 &  & \cellcolor[rgb]{0.993248,0.906157,0.143936} 1 \\
    schemathesis:Default & \cellcolor[rgb]{0.993248,0.906157,0.143936} 1 &  &  & \cellcolor[rgb]{0.993248,0.906157,0.143936} 1 & \cellcolor[rgb]{0.993248,0.906157,0.143936} 1 & \cellcolor[rgb]{0.993248,0.906157,0.143936} 1 &  & \cellcolor[rgb]{0.993248,0.906157,0.143936} 1 & \cellcolor[rgb]{0.993248,0.906157,0.143936} 1 &  & \cellcolor[rgb]{0.993248,0.906157,0.143936} 1 & \cellcolor[rgb]{0.993248,0.906157,0.143936} 1 & \cellcolor[rgb]{0.993248,0.906157,0.143936} 1 &  &  & \cellcolor[rgb]{0.993248,0.906157,0.143936} 1 \\
    schemathesis:Negative & \cellcolor[rgb]{0.993248,0.906157,0.143936} 1 &  &  & \cellcolor[rgb]{0.993248,0.906157,0.143936} 1 & \cellcolor[rgb]{0.993248,0.906157,0.143936} 1 & \cellcolor[rgb]{0.993248,0.906157,0.143936} 1 &  & \cellcolor[rgb]{0.993248,0.906157,0.143936} 1 & \cellcolor[rgb]{0.993248,0.906157,0.143936} 1 &  & \cellcolor[rgb]{0.993248,0.906157,0.143936} 1 & \cellcolor[rgb]{0.993248,0.906157,0.143936} 1 & \cellcolor[rgb]{0.993248,0.906157,0.143936} 1 &  &  & \cellcolor[rgb]{0.993248,0.906157,0.143936} 1 \\
    swagger\_fuzzer &  &  &  &  &  &  &  & \cellcolor[rgb]{0.993248,0.906157,0.143936} 1 &  &  &  &  &  &  &  &  \\
    tnt\_fuzzer &  &  &  &  &  &  &  &  &  &  &  &  &  &  &  & \cellcolor[rgb]{0.993248,0.906157,0.143936} 1 \\
    
    \midrule
    \multicolumn{17}{l}{
        \parbox{0.95\linewidth}{
            \caption{Mean number of reports per unique defect, of those reported in a given run. Hypothesis' shrinking allows Schemathesis to report a single easy-to-understand example for each defect, easing triage.}
            \vspace{-0.8em}
            \label{table:n_reports}
    }} \\

    \bottomrule
  \end{tabular}
\end{table*}

\begin{figure}
\centering
\includegraphics[width=\columnwidth]{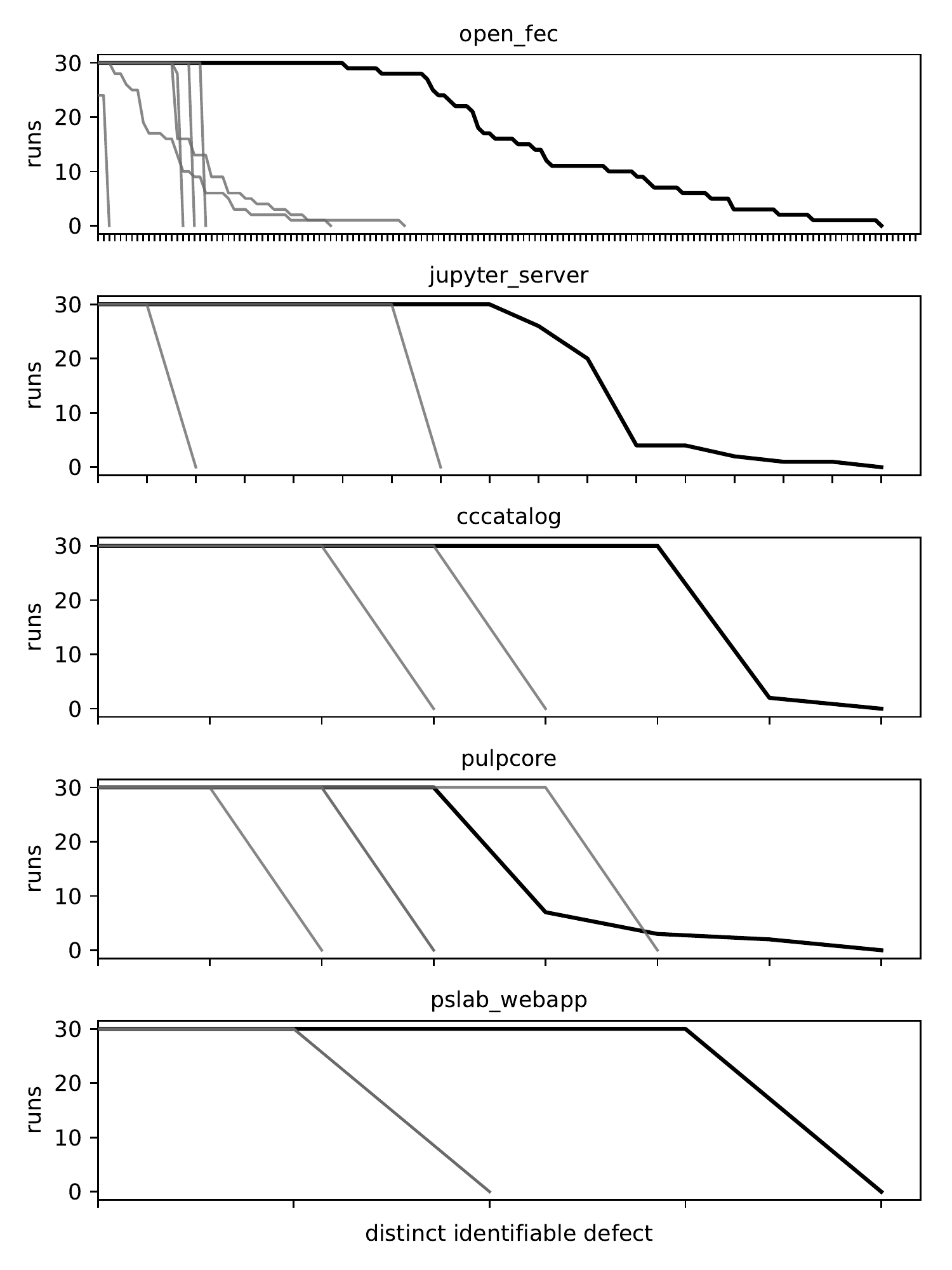}
\caption{
Sorting unique defects by the number of runs detecting them, we see that
Schemathesis (black) is less consistent but discovers more than other fuzzers.
Defect IDs are consistent between runs but not between fuzzers.}
\label{fig:consistency}
\end{figure}

In all-checks mode, Schemathesis reports a total of 755 bugs across
fourteen out of our sixteen targets, including 111 \texttt{HTTP 500}
responses, 436 unexpected status codes, 52 non-schema-conforming
responses, and 152 responses with a wrong or missing content type.
Schemathesis finds $1.4\times$ to $4.5\times$ more defects
than the respectively second-best fuzzer for each target, and is the
\emph{only} fuzzer to find defects in four targets.

Schemathesis was the only one of the tools we evaluated to handle all--or
more than two-thirds of--our targets without a fatal internal error.
Surprisingly, this appears unrelated to our pairing of OpenAPI 3 targets
with fuzzers which do not claim to support OpenAPI 3.

Table \ref{table:uniq_failures} shows a summary of identified defects
by fuzzer and target, some of which might be resolved by making the schema
more permissive, particularly the unexpected status codes. We argue that
such mismatches between specified and actual behaviour are still
reasonably described as bugs - the schema being no less important
than the implementation of the service.  The issues listed in Table
\ref{table:in-the-wild} indicates that at least some of our users agree.

To support clear comparisons to fuzzers which check fewer semantic properties,
Table \ref{table:uniq_500s} shows only \texttt{HTTP 500} internal server errors.
We see a rich understanding of application semantics, including over request
sequences, as a key contribution of our research.  Nonetheless, Schemathesis
detected more unique errors than the respectively second-best fuzzer for
each target.

Property-based testing workflows also outperform fuzzing when it
comes to actionable reporting.  Table \ref{table:n_reports} shows the mean
number of reports per unique defect, averaged over each independent run.
Schemathesis reports a single minimal triggering input or action-sequence
for each, with the exception of a few cases where our evaluation harness
uses additional information from server logs to deduplicate internal errors.

Figure \ref{fig:consistency} shows the consistency of defect-detection
between runs.  For most tools, detection is binary: if they do not discover
a defect on the first run, they are unlikely to ever do so.  APIFuzzer, Cats,
and Schemathesis are less consistent for all but the easiest defects.

In Schemathesis, we attribute this effect to the property-based testing workflow!
Because the user is expected to fix each failing test by changing either
the service or the test harness, Hypothesis stops looking shortly after
finding the first failing input.  When there are multiple defects which
may be discovered in different orders, this early stopping also reduces
consistency compared to long-running fuzzing workflows
(Table \ref{table:durations}).

If inconsistency is driven by low detection probabilities instead of early
stopping, this is also good news for users:  they can simply run these
fuzzers for longer and discover more defects---and in the absence of a
run reporting \textit{zero} defects, they are likely to do so.

\section{Directions for future research}

\paragraph{Expand and share benchmark suite.}  Our evaluation suite is
designed to be reused, and will be maintained as an open source project
by the Schemathesis developers.  Adding and updating REST and GraphQL fuzzers
\footnote{e.g. RESTest \cite{RESTest}, EvoMaster \cite{EvoMaster},
and Karlsson et al \cite{karlsson2020automatic}}
and targets, improving our understanding of the
`ground truth' by manually identifying defects or better automated tools
for triage, and collaboration among fuzzer developers would all
be valuable contributions.

\paragraph{Build defect analysis into fuzzers}
A single \texttt{HTTP 500} internal error should be reported once, not once-per-triggering-endpoint---and automatic defect analysis works well
enough in our evaluation that we think this is within reach, at least when
running in-process or with an existing monitoring solution\footnote{
   respectively e.g. Schemathesis ASGI/WSGI support, or Sentry
}.

\paragraph{Support the long tail of schema features.}
Schemathesis has gone further down this path than previous fuzzers, and
outperforms accordingly.  What further gains are locked behind support
for rare features such as ECMA-specific regex syntax, non-unrollable
recursive references, or niche content-types?

\paragraph{Investigate code-coverage-guided fuzzing for web APIs.}
While coverage-guided fuzzing has been highly successful for native
code, to date it has been impractical to evaluate the value of coverage
feedback for web API fuzzing.  Schemathesis' support for ASGI/WSGI in-process
fuzzing, and EvoMaster's JVM instrumentation, allow measuring the value of
coverage feedback\footnote{
as distinct from the client-observable notions of behavioural
coverage that can be explored using e.g. Schemathesis targets.}
in present systems.

\paragraph{Improved request-sequence testing without ``links''}
Schemathesis is often limited by the absence of OpenAPI ``links'' describing
how data from one response can be used to make further requests.  Investigating
\cite{RESTTESTGEN,REST-ler}-style heuristics or ways to learn links from
traffic records could substantially improve performance.

\paragraph{Recommend schema improvements}
Tools could suggest improvements to API schemas, whether to resolve basic
type confusions or refine their semantics.  This would be particularly
valuable in combination with learned links between endpoints.

\paragraph{Embed property checks into web frameworks}
Web frameworks which dynamically generate the application schema from code
could also check conformance with the schema at runtime.  Even if limited
to a debug mode for performance, such automatic test oracles could improve
the effectiveness of property-na\"ive fuzzers and other forms of testing

\section{Conclusion}

Building on mature and versatile property-based testing tools makes
Schemathesis easily adaptable to specific services or workflows and
remarkably effective. Working with Hypothesis offers a integrated and
growing suite of useful techniques which we would not otherwise have
implemented.

Schemathesis is the only fuzzer in our comprehensive evaluation to
handle every real-world schema and web service, and consistently reports
more defects than the previous state-of-the-art.

We hope that future work on web API fuzzing will reuse---and perhaps
extend---our summary of testable properties and our evaluation corpus
of tools and services.

\balance
\bibliographystyle{ACM-Reference-Format}
\bibliography{references}

\end{document}